\documentclass[pre,showpacs,preprint]{revtex4}

\usepackage{amsmath}

\usepackage{graphicx}
\usepackage{dcolumn}
\usepackage{bm}

\begin{document}

\title{Memory effects in the relaxation of a confined granular gas}
\author{J. Javier Brey, M.I. Garc\'{\i}a de Soria, P. Maynar, and V. Buz\'{o}n}
\affiliation{F\'{\i}sica Te\'{o}rica, Universidad de Sevilla,
Apartado de Correos 1065, E-41080, Sevilla, Spain}
\date{\today }

\begin{abstract}
The accuracy of a model to describe the horizontal dynamics of a confined quasi-two-dimensional system of inelastic hard spheres is discussed by comparing its predictions for the relaxation of the temperature in an homogenous system with molecular dynamics simulation results for the original system. A reasonably good agreement is found. Next, the model is used to investigate the peculiarities of the nonlinear evolution of the temperature  when the parameter controlling the energy injection is instantaneously changed while the system was relaxing. This can be considered as a non-equilibrium generalization of the Kovacs effect. It is shown that, in the low density limit, the effect can be accurately described by using a simple kinetic theory based on the first Sonine approximation for the one-particle distribution function. Some possible experimental  implications are indicated.
\end{abstract}


\maketitle

\section{Introduction}
\label{s1}
Granular gases \cite{Go03} are composed of macroscopic particles, grains, which do not conserve the kinetic energy when they collide. As such, there is no equilibrium state, since an isolated undriven system will cool continuously. To reach and maintain a steady state, a permanent injection of energy is needed. Typically, this is done by injecting energy through the boundaries or by means of external fields acting on the bulk of the system. As a consequence, the system develops large inhomogeneities, often associated to the presence of instabilities. A driven homogeneous steady state can be obtained if the grains are submitted to a (fictitious)  external noise force, usually referred to as a stochastic thermostat \cite{vNyE98,PLMyV99,PTvNyE01}. Nevertheless, the possible relationship of this kind of forces with real experimental situations has not been discussed in detail, to the best of our knowledge. 

In the last years a quite interesting alternative, overcoming the above difficulty, has been addressed. The idea is to consider a granular gas confined to a quasi-two-dimensional geometry by placing it between two large  parallel horizontal plates separated a distance smaller than two grain diameters, so that the system is actually a granular monolayer \cite{OyU05,RPGRSCyM11}.  To keep the system fluidized and render a steady state possible, it is mechanically vibrated in the vertical direction.  The two-dimensional dynamics of the projection of the grains motion on an horizontal plane is considered. It has been found experimentally that the behavior of the projected system resembles that of a two-dimensional fluid. Moreover, it remains homogeneous over a wide range of values of the parameters controlling the experiment, eventually reaching a steady homogeneous state. On the other hand, when varying the average density and the intensity of the vibration, the system exhibits a a series of phase transitions \cite{OyU05,CMyS12,PCyR09,RPGRSCyM11}.

A common paradigm for granular gases, employed in many theories and simulations, is that of a collection of uniform smooth inelastic hard spheres or disks, whose collisions are characterized by a constant, velocity independent, coefficient of normal restitution \cite{ByP04}. Although this model has several shortcomings, it enables to make the relevance of inelasticity clear in the simplest possible scenario.  Very recently \cite{BRyS13}, this model has been modified trying to mimic the horizontal dynamics of a vibrated fluid of inelastic hard spheres confined to a quasi-two-dimensional geometry, i.e. the kind of system described above. To do so, the inelastic collision rule for hard disks has been changed in order to account for the mechanism by which the kinetic energy given to the vertical motion of the particles by the vibrating walls is transferred to the horizontal degrees of freedom through the collisions between particles. Since the model is formulated in terms of a deterministic instantaneous collision rule, the methods of non-equilibrium statistical mechanics and kinetic theory developed for inelastic hard spheres and disks \cite{BDyS97,vNEyB98} can be easily extended to it \cite{BGMyB13}. In this way, it has been shown that it admits a solution describing a homogeneous steady state, similarly to what happens in the experiments \cite{BGMyB13}.

A first relevant question is to somehow evaluate the accuracy of the model to describe the original quasi-two-dimensional vibrated system and, if possible, to compare it with the stochastic thermostat model mentioned at the beginning. Some information about this issue will be provided here by analyzing  the relaxation of the temperature towards its steady value as predicted by both models, and the results obtained by means of molecular dynamics (MD) simulations of the the original three-dimensional confined system.

As a direct implication  of the existence of a mechanism of energy injection, there is homogeneous hydrodynamics in the collisional model proposed by Brito {\em et al.} \cite{BRyS13}, i.e. there is a time scale over which the time evolution of the (granular) temperature in an homogeneous system obeys a macroscopic differential equation whose solution only depends on the initial temperature \cite{BMGyB14}. In the context of kinetic theory, the existence of homogeneous hydrodynamics follows from the one-particle distribution function of the system being {\em normal},  meaning that all its time dependence occurs through the temperature \cite{RydL77,DyB11}. Therefore, two homogeneous hydrodynamic states of a given system are the same state if both have the same temperature. The above is true not only at the macroscopic level, but also when the system is described by means of kinetic theory or non-equilibrium statistical mechanics using its distribution function. Of course, there is the possibility of different homogeneous states having the same temperature, but only one of them, at the most, can be described by hydrodynamics. The above comments apply, in particular, to the steady state, since it  is the hydrodynamic state reached by the system in the long time limit. 

Consider a confined quasi-two-dimensional  system which is in the homogeneous stationary state with a temperature $T_{1}$. Of course, this refers to the projected two-dimensional  system. Then, the vibration intensity is instantaneously decreased so that the new stationary temperature $T_{2}$ is lower, and the system will cool towards it. At a later time $t_{0}$, the vibration intensity is again modified, now in such a way that  the instantaneous value of the temperature $T(t_{0})$ be equal to the steady temperature corresponding to the new vibration intensity. Although the temperature of the system has its long time steady value, it is not expected to remain constant, but to deviate from this steady value, going through a maximum or a minimum, and finally tending  to  stationarity. The reason to expect this behavior  is that, although the system has its steady temperature, it is not in the steady state. Even more, it can not be in a hydrodynamic state for the reasons mentioned above. As a consequence, the macroscopic equation for the evolution of the temperature  can not be used to describe its evolution following the change of the vibration intensity at  $t_{0}$. The experiment just described as well as  the predicted behavior of the system, can be considered as reminiscent of the Kovacs memory effect occurring in the relaxation of molecular systems towards equilibrium \cite{Ko63}, that has been extensively studied in the last years \cite{ByH02,MyS04,AAyN08,PyB10,DyH11}. Nevertheless, it is important to realize that in the above process we are dealing with an intrinsic non-equilibrium system trying to approach its steady state, while the original Kovacs effect refers to the relaxation towards equilibrium. Moreover, while Kovacs carried out an instantaneous quench of the system and studied the evolution of a macroscopic quantity, namely the volume of the system, here what is instantaneously changed is the external parameter controlling the energy injection, and it is the time evolution of the temperature what is followed.

One of the aims of this paper is to investigate the presence of a Kovacs-like effect in the collisional model for a confined granular gas. It will be shown that this is the case and that, for a low density granular gas,  the effect is well captured by  kinetic theory at the level of the (inelastic) Boltzmann equation in the first Sonine approximation. Actually, the effect shows up both when the characteristic speed involved in the formulation of the model of the model is modified both to increase and to decrease the value of the steady temperature and, therefore, leads to a heating and cooling process, respectively.

This paper is organized as follows. In the next section, the stochastic thermostat model \cite{vNyE98,PTvNyE01} and the collisional model  \cite{BRyS13,BGMyB13} leading to homogeneous steady  states of a system of inelastic hard disks are shortly reviewed. The macroscopic equations for the evolution of the granular temperature and the expressions for its steady value in both models are reminded.  By fitting the steady temperatures predicted by the models to the results obtained for the horizontal motion of a confined quasi-two-dimensional system of inelastic hard spheres under vibration by MD simulations, the phenomenological parameter appearing in each of the models is determined, as a function of the vibration intensity. Then, these values are used into the equations for the evolution of the temperature predicted by each model, and their solutions  are compared with MD simulation results, without any new fitting parameter.  The agreement of the collisional model is definitely better than that of the stochastic thermostat model. In Sec. \ref{s3}, the steady velocity distribution is considered. The MD results indicate that the shape of the distribution is very close to a Gaussian. Therefore, the first Sonine approximation has been considered, and the comparison of the two models with the simulation data has been carried out by using the fourth moment of the velocity distribution, which is closely related to the coefficient of the first Sonine correction term. Although both, the stochastic and the collisional models, accurately predict that the Sonine coefficient is very small, they fail to describe its qualitative dependence with the coefficient of normal restitution of the particles. 

In Sec. \ref{s4}, an ideal experiment in which the characteristic speed of the collisional model is suddenly changed is described, and analyzed  in the low density limit by using kinetic theory. Its predictions for the nonlinear time evolution of the temperature after the jump are compared with numerical results obtained using the direct simulation Monte Carlo method. A reasonable good agreement is found. Finally, the last sections contains some summarizing remarks as well as some possible experimental implications of the results reported in the paper.

\section{Two models of a vibrated confined granular  gas}
\label{s2}	
The system considered is a vibrated granular gas confined to a quasi-two-dimensional geometry. The grains are smooth, inelastic hard spheres of mass $m$ and diameter $\sigma$. Collisions between particles are characterized by a constant, velocity independent, coefficient of normal restitution $\alpha$, defined in the interval $0 < \alpha \leq 1$. Therefore, when two particles, $i$ and $j$, collide their velocities are instantaneously modified according with
 \begin{equation}
\label{2.1}
{\bm v}_{i} \rightarrow {\bm v}^{\prime}_{i} = {\bm v}_{i}- \frac{1+ \alpha}{2}  {\bm v}_{ij} \cdot \widehat{\bm \sigma}  \widehat{\bm \sigma},
\end{equation}
\begin{equation}
\label{2.2}
{\bm v}_{j} \rightarrow {\bm v}^{\prime}_{j} = {\bm v}_{j}+ \frac{1+ \alpha}{2}  {\bm v}_{ij} \cdot \widehat{\bm \sigma} \widehat{\bm \sigma},
\end{equation}
where ${\bm v}_{ij} \equiv {\bm v}_{i}-{\bm v}_{j}$ is the relative velocity prior collision and $ \widehat{\bm \sigma}$ is a unit vector directed from the center of particle $j$ to the center of particle $i$ at contact. 

The system is confined by means of two parallel plates separated a distance $h$, $\sigma <h< 2 \sigma$. The bottom plate is vibrating, and for the sake of simplicity, the limit of small amplitude and very high frequency will be supposed. Then the bottom wall moves in a sawtooth way, and every colliding particle  finds it at a fixed position and moving with a constant speed $v_{b}$ directed upwards \cite{McyB97}. The upper wall is at rest. Collisions of the particles with both walls are elastic. As the particles collide with the bottom wall, energy is transferred to the vertical degree of freedom of the grains. Afterwards, this energy is transferred to the horizontal degrees of freedom of the grains in the collisions between them. The interest here is on the behavior of the system when observed from above and the two-dimensional dynamics in the horizontal plane is analyzed. No effect of gravity is considered. The system has been  designed trying to mimic some experimental setups reported in the literature \cite{OyU99,CMyS12}. Also, a similar confined granular system, but without the top lid has been investigated \cite{PGGSyV12}.  In both cases, one of the pursued goals  is to avoid the large inhomogeneities usually developed in the bulk of a granular gas when energy is injected  inhomogeneously through the boundaries \cite{ByC96,BRyM01}. In the confined quasi-two-dimensional granular gas considered here, homogeneous steady states are possible because the energy injection occurs throughout the whole system \cite{OyU98,PMEyU04,Metal05,CCDHMRyV08}

Molecular dynamics (MD) simulations of the above system have been carried out by using the event-driven method \cite{AyT87,PyS05}. Periodic boundary conditions have been used in the horizontal plane. In the initial state, grains were homogeneosuly distributed in the horizontal plane equidistant from both walls, and the initial velocities were assigned accordingly with a  Gaussian distribution of zero average. Details of these initial configurations are always lost in a few collisions per particle and, if the two-dimensional dynamics is considered, the system reaches a time-dependent homogeneous state. In the long time limit,  the system eventually remains in a homogeneous steady state, although when varying the solid fraction and the vibration amplitude, the system exhibits a solid-liquid transition, with different possible solid phases \cite{OyU98,PMEyU04,Metal05}. For fixed values of the geometrical parameters defining the particles and the system, the density, and the coefficient of normal restitution, the steady temperature depends on the velocity $v_{b}$ of the vibrating wall.

 Two models have been proposed in the literature that allows for a homogeneous steady state of a granular gas. In the first one, the granular medium is subjected to a bulk injection of energy by means of a stochastic force \cite{vNyE98}. The model was formulated without having any particular application in mind. 
On the other hand, the second model was designed as a collisional model trying to mimic the effective two-dimensional dynamics of the confined quasi-two-dimensional granular gas \cite{BRyS13}. Next, both models will be shortly summarized, focussing on the subject of interest here, namely their predictions for the homogeneous macroscopic evolution of the two-dimensional granular gas.

\subsection{The stochastic thermostat}
\begin{figure}
 \includegraphics[scale=0.4,angle=0]{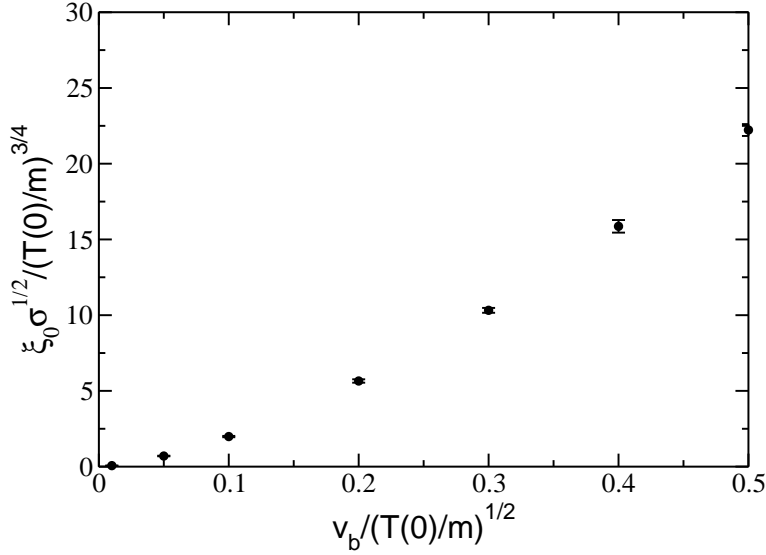}
\caption{Noise strength  $\xi_{0}$ of the stochastic thermostat model for the two-dimensional dynamics as a function of the velocity $v_{b}$ of the vibrating wall in  the three-dimensional system, obtained in the way discussed in the main text. Dimensionless  quantities have been defined as indicated in the respective labels, with $T(0)$ being the initial temperature of the (three-dimensional) gas. The coefficient of normal restitution is $\alpha=0.8$ and the density of the gas $n=0.02 \sigma^{-3}$.   \label{f1}}
\end{figure}

In the stochastic thermostat model, energy is uniformly supplied to the system by means of a white noise force acting independently on each grain \cite{vNyE98,PTvNyE01,GyM02,VPBTyvW06,GMyT09,MGyT09}, so that the one-particle velocity distribution, $f({\bm v},t)$, of a dilute homogeneous gas of inelastic hard disks obeys the Boltzmann-Fokker-Planck equation
\begin{eqnarray}
\label{2.3}
\frac{\partial}{\partial t}\, f({\bm v}_{1},t) & = & \sigma \int d {\bm v}_{2} \int d \widehat{\bm \sigma}\,  \theta \left( {\bm v}_{12} \cdot \widehat{\bm \sigma} \right) {\bm v}_{12} \cdot \widehat{\bm \sigma} \left[ \alpha^{2} b_{\bm \sigma}^{-1} (1,2) -1 \right] f({\bm v}_{1},t) f({\bm v}_{2},t) \nonumber \\
&& + \frac{\xi_{0}^{2}}{2}\, \frac{\partial^{2}}{\partial {\bm v}^{2}}\, f({\bm v}_{1},t).
\end{eqnarray}
Here the operator $b_{\bm \sigma}^{-1}(1,2)$ changes the velocities ${\bm v}_{1}$ and ${\bm v}_{2}$ to its right into the precollisonal values corresponding to the collision rule given in Eq.\ (\ref{2.1}) and (\ref{2.2}), i.e.
\begin{equation}
\label{2.4} b_{\bm \sigma}^{-1} (1,2) {\bm v}_{1} = {\bm v}_{1} - \frac{1+ \alpha}{2 \alpha}\,  {\bm v}_{12} \cdot \widehat{\bm \sigma}  \widehat{\bm \sigma},
\end{equation}
\begin{equation}
\label{2.5} b_{\bm \sigma}^{-1} (1,2){\bm v}_{2} = {\bm v}_{2} + \frac{1+ \alpha}{2 \alpha}\, {\bm v}_{12} \cdot \widehat{\bm \sigma}  \widehat{\bm \sigma}  .
\end{equation}
The quantity $\xi_{0}$ measures the magnitud of the fluctuating force and it must be considered as a parameter defining the model. Approximating the distribution function by a Gaussian, an evolution equation for the temperature $T(t)$ is obtained by taking moments in the above Boltzmann equation, 
\begin{equation}
\label{2.6}
\frac{\partial T}{\partial t} = - \Gamma + m \xi_{0}^{2},
\end{equation}
where
\begin{equation}
\label{2.7}
\Gamma \equiv \frac{(1-\alpha^{2}) \omega T}{2}
\end{equation}
and
\begin{equation}
\label{2.8}
\omega \equiv 2 \pi n \sigma \left( \frac{T}{\pi m} \right)^{2/3} ,
\end{equation}
$n$ being the number density of disks. Therefore, the steady temperature $T_{st}$ reached by the system in the long time limit is given by
\begin{equation}
\label{2.9}
T_{st}= m \left[ \frac{\xi_{0}^{2}}{ \pi^{1/2} (1-\alpha^{2}) n \sigma} \right]^{2/3}.
\end{equation}
This expression has been used to determine the unknown parameter $\xi_{0}$ of the model as a function of the velocity $v_{b}$ of the vibrating wall in the original quasi-two-dimensional system. The steady temperature of the effective two-dimensional gas in the latter has been measured by means of MD simulations as a function of $v_{b} $. Equation (\ref{2.9}) provides the value predicted by the stochastic model as a function of the noise amplitud $\xi_{0}$. It is then trivial to fit $\xi_{0}$ as a function of $v_{b}$, for fixed values of all the other parameters, so that Eq.\ (\ref{2.9}) yields the dependence of $T_{st}$ on $v_{b}$ measured in the MD simulations. In Fig. \ref{f1}, the results obtained for a system of $N=500$ inelastic hard spheres with a coefficient of normal restitution $\alpha=0.8$ are shown. The unit cell is a parallelepipedal of basis $L\times L$, with $L=129 \sigma$,  and height $h=1.5 \sigma$. Then, the three-dimensional number density is $n=0.02 \sigma^{-3}$, which corresponds to a dilute granular gas. The two-dimensional number density entering in Eq. (\ref{2.9}) has been computed by considering $N$  disks of diamater $\sigma$ inside the square of size $L \times L$. 

\subsection{The collisional model}

The idea in the collisional model is to replace the actual three-dimensional  collision rule, Eqs. (\ref{2.1}) and (\ref{2.2}), by an effective two-dimensional one to describe  the dynamics in the horizontal plane  \cite{BRyS13}. Then, a system of hard disks is considered, in which collisions change the velocities according to
\begin{equation}
\label{2.10}
{\bm v}_{i} \rightarrow {\bm v}^{\prime}_{i} = {\bm v}_{i}- \frac{1+ \alpha}{2}  {\bm v}_{ij} \cdot \widehat{\bm \sigma}  \widehat{\bm \sigma} + \Delta \widehat{\bm \sigma},
\end{equation}
\begin{equation}
\label{2.11}
{\bm v}_{j} \rightarrow {\bm v}^{\prime}_{j} = {\bm v}_{j}+ \frac{1+ \alpha}{2}  {\bm v}_{ij} \cdot \widehat{\bm \sigma} \widehat{\bm \sigma} - \Delta \widehat{\bm \sigma}.
\end{equation}
Here $\Delta$ is some positive characteristic speed. It tries to describe in an effective way the mechanism for which energy is injected into the horizontal degrees of freedom of the inelastic hard spheres. 

\begin{figure}
\includegraphics[scale=0.4,angle=0]{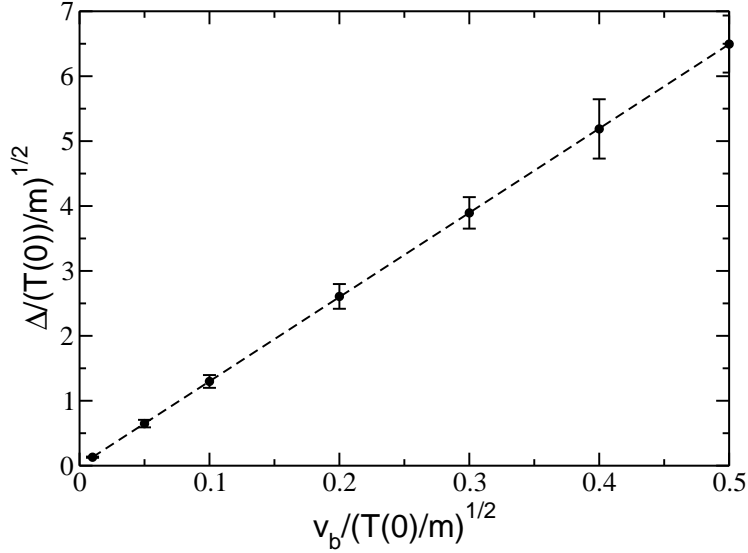}
\caption{ Characteristic speed $\Delta$ of the collisional  model for the two-dimensional dynamics as a function of the velocity $v_{b}$ of the vibrating wall in  the three-dimensional system, obtained in the way discussed in the main text. Dimensionless quantities have been defined as indicated in the respective labels, with $T(0)$ being the initial temperature of the (three-dimensional) gas. The coefficient of normal restitution is $\alpha=0.8$ and the density of the gas $n=0.02 \sigma^{-3}$. The straight dashed line is just a guide for the eye.  \label{f2}}
\end{figure}

In the low density limit, the dynamics of the system can be accurately described by an inlelastic Boltzmann equation \cite{BGMyB13}. From it, an evolution equation for the temperature in homogeneous situations can be derived. It reads \cite{BMGyB14}:
\begin{equation}
\label{2.12}
\frac{\partial T(t)}{\partial t} = - \zeta_{H}(t) T(t),
\end{equation}
with the homogeneous rate of change  of the temperature $\zeta_{H}(t)$ given by
\begin{equation}
\label{2.13}
\zeta_{H}(t)= ( 2\pi)^{1/2}n \sigma v_{0}(t) \left[ \frac{1 - \alpha^{2}}{2} \left( 1+ \frac{3 a_{2}}{16} \right) - \alpha \left( \frac{\pi}{2} \right) ^{1/2} \Delta^{*} - \left( 1- \frac{a_{2}}{16} \right) \Delta^{*2} \right],
\end{equation}
where $v_{0}(t) \equiv \left( 2T(t)/m \right)^{1/2}$ is a thermal velocity and $\Delta^{*} \equiv \Delta / v_{0}(t)$. Upon deriving the above expression, the first Sonine approximation \cite{RydL77} for the one-particle distribution function has been assumed,
\begin{equation}
\label{2.14}
f({\bm v},t) \approx n v_{0}(t)^{-2} \pi ^{-1} e^{-c^{2}} \left[ 1 + a_{2} (t) S^{(2)} (c^{2}) \right].
\end{equation}
 In this expression, 
 \begin{equation}
 \label{2.15}
 {\bm c} \equiv \frac{\bm v}{v_{0}(t)}
 \end{equation}
 and
 \begin{equation}
 \label{2.16}
 S^{(2)}(c^{2}) \equiv 1-2 c^{2} + \frac{c^{4}}{2}.
 \end{equation}It is worth to stress that Eqs. (\ref{2.12}) and (\ref{2.13}) hold as long as the distribution function can be approximated by  (\ref{2.14}). The expression of the Sonine coefficient $a_{2}(t)$ in the hydrodynamic regime has been determined in ref. \cite{BMGyB14} and it will be considered in Sec. \ref{s4}. For the moment being, an accurate enough result for the steady temperature is obtained by using the Gaussian approximation for the distribution function, i.e. by putting $a_{2}=0$ in Eq. (\ref{2.13}). In this case, the steady temperature predicted by the collisional model is \cite{BRyS13,BGMyB13}
 \begin{equation}
 \label{2.17}
 T_{st}= \frac{\pi m \alpha^{2}}{4(1-\alpha^{2})^{2}} \left[ 1+ \sqrt{ 1 + \frac{4(1-\alpha^{2})}{\pi \alpha^{2}}} \right]^{2} \Delta^{2}.
 \end{equation}
 Now the same procedure as for the stochastic thermostat model can be used to determine the value of the characteristic speed $\Delta$ from the MD simutation results in the present model. In Fig. \ref{f2}, the characteristic speed $\Delta$ is plotted as a function of the velocity of the vibrating wall $v_{b}$ for the same system as considered in Fig. \ref{f1}. A quite perfect linear relationship is observed, consistently with dimensional analysis.

\subsection{Time evolution of the temperature}

\begin{figure}
\includegraphics[scale=0.4,angle=0]{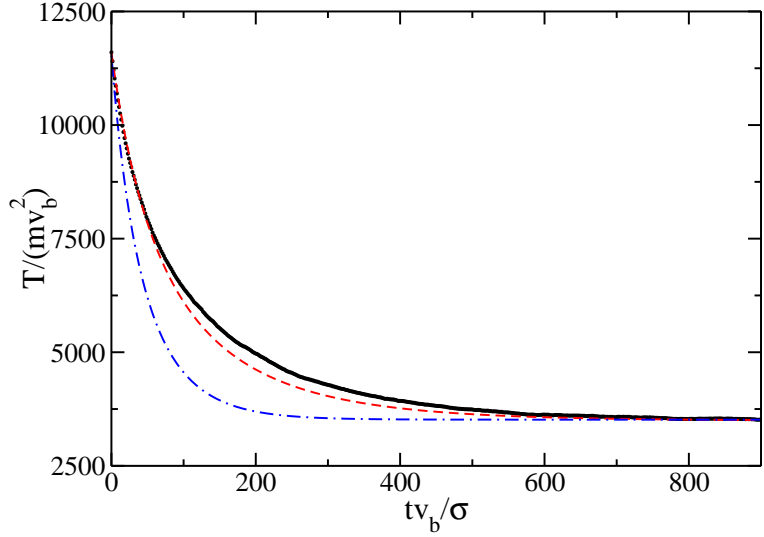}
\caption{(Color online) Relaxation of the temperature in a system with a coefficient of normal restitution  $\alpha=0.8$,  height $h=1.5 \sigma$, and three-dimensional density $n= 0.02 \sigma^{-3}$. The velocity of the vibrating wall is $v_{b}=0.01 \left( T(0)/m \right)^{1/2}$. Both temperature and time are measured in the dimensionless units indicated in the labels.  The solid line is the result from MD simulations, while  the dot-dashed (blue) line and the (red) dashed line are the theoretical predictions from the stochastic and collisional models, respectively.  The used values of $\xi_{0}$ and $\Delta$  are from Figs. \ref{f1} and \ref{f2}. \label{f3}}
\end{figure}

\begin{figure}
\includegraphics[scale=0.4,angle=0]{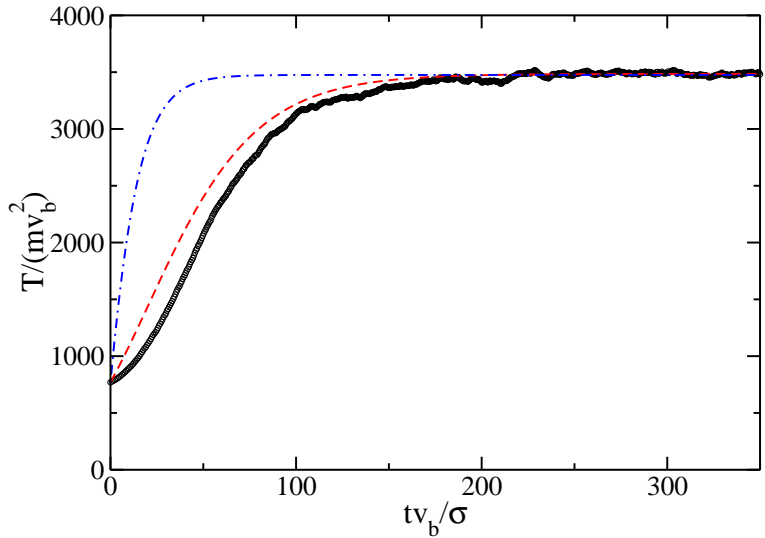}
\caption{ (Color online) The same as described in the legend of Fig. \protect{\ref {f3}}, but now for a system with a vibrating wall with a velocity  $v_{b}=0.05 \left( T(0)/m \right)^{1/2} $. \label{f4}}
\end{figure}

Once the values of the parameters $\xi_{0}$ and $\Delta$ defining each of the two models have been determined by fitting the steady temperature measured in the MD simulations of the confined granular gas, those values can be used to test the accuracy of the predictions for the time evolution of the temperature towards its steady value. In Figs. \ref{f3} and \ref{f4} the results obtained for two different values of the vibration velocity $v_{b}= 0.01 \left(T(0) / m \right)^{1/2}$ and $v_{b}=0.05  \left(T(0) / m \right)^{1/2}$, respectively,  are shown. The density, the height $h$,  and the coefficient of normal restitution $\alpha$ are the same as in Figs.\ \ref{f1} and \ref{f2}. In the first case the system monotonically cools towards the steady state, while in the latter the system heats while approaching stationarity. The solid lines are simulation data, the dot-dashed lines are the theoretical predictions obtained with the stochastic model, and the dashed lines are the predictions of the collisional model. It is seen that the prediction of the collisional model is much better than that of the stochastic model, in which the initial rate of variation of the temperature is much larger than in the MD simulations. Similar results have been  obtained for other values of the vibration velocity $v_{b}$, although the accuracy of the theoretical predictions gets worse as the velocity increases.

\section{The steady velocity distribution}
\label{s3}
The steady  one-particle velocity distribution in the horizontal plane, $f_{st}({\bm v})$, has also been measured in the MD simulations of the  granular gas confined  between two paralell plates. Taking into account the isotropy of the system in the horizontal plane, results will be presented in the following for the marginal velocity distribution $\varphi _{x} (c_{x})$ defined by
\begin{equation}
\label{3.1}
\varphi_{x}(c_{x}) \equiv \int_{-\infty}^{+ \infty} d c_{y}\, \varphi (c),
\end{equation}
where $ \varphi(c)$ is the scaled velocity distribution of the dimensionless velocity ${\bm c} \equiv  {\bm v}/v_{0,st}$, with $v_{0,st} \equiv \left( 2 T_{st}/m \right)^{1/2}$,
\begin{equation}
\label{3.2}
f_{st}({\bm v}) \equiv n v_{0,st}^{-2} \varphi (c).
\end{equation}
The simulation results indicate that the shape of the velocity distribution is very similar to a Gaussian. This leads to try to describe it by means of a truncated Sonine expansion. To lowest order, it has the form \cite{RydL77}
\begin{equation}
\label{3.3}
\varphi_{x}(c_{x}) \approx  \pi^{-1/2} e^{-c_{x}^{2}} \left[ 1 + \frac{a_{2,st}}{2} \left( \frac{3}{4}-3 c_{x}^{2} + c_{x}^{4} \right) \right].
\end{equation}
The Sonine coefficient $a_{2,st}$ is related with the forth moment of the velocity distribution by
\begin{equation}
\label{3.4}
a_{2,st} \equiv \frac{<c^{4}>}{2}\, -1,
\end{equation}
with
\begin{equation}
\label{3.5}
<c^{4}> \equiv \int d{\bm c}\, c^{4} \varphi (c).
\end{equation}

\begin{figure}
\includegraphics[scale=0.4,angle=0]{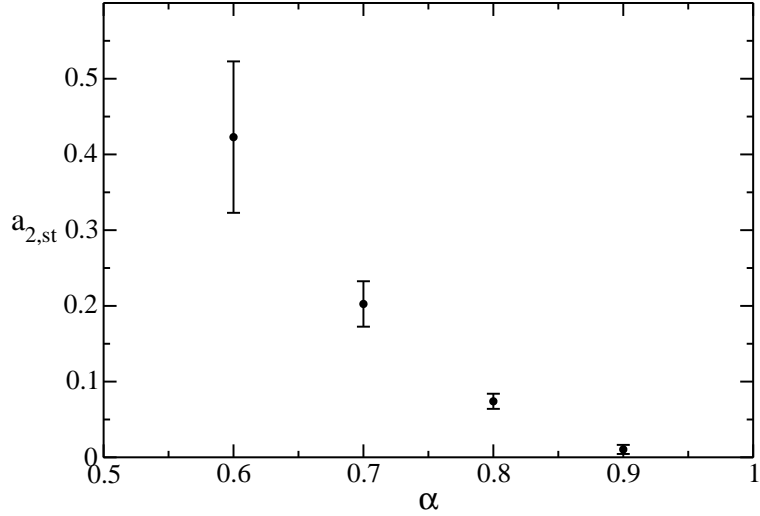}
\caption{ MD simulation results for the Sonine coefficient $a_{2,st}$ of the steady velocity distribution function in the horizontal plane of the quasi-two-dimensional granular gas  as a function of the coefficient or normal restitution $\alpha$. The velocity of the vibrating wall is $v_{b}=  0.01 \left( T(0)/m \right)^{1/2}$, the density of the three-dimensional gas is $n=0.02 \sigma^{-3}$, and the distance between the two confining walls is $h= 1.5 \sigma$. \label{f5}}
\end{figure}

In the simulation results being reported, $a_{2,st}$ has been measured using Eq. (\ref{3.4}). In Fig. \ref{f5}, the Sonine  coefficient is plotted as a function of the coefficient of normal restitution for a system of $500$ inelastic hard spheres  with $v_{b}= 0.01 \left( T(0)/m \right)^{1/2}$, $n= 0.02\sigma^{-3}$, and $h=1.5 \sigma$. Two features of the simulation data must be emphasized. First, $a_{2,st}$ is positive in all the range of $\alpha$ investigated. Second, it is very small for moderate and weak inelasticity, say $\alpha \agt 0.8$, while it grows rather fast for stronger inelasticities. It is interesting to compare this behavior with the results obtained from the two models discussed in the previous section. The collisional model predicts \cite{BGMyB13} that the coefficient is negative and very small ( $\agt-  0.12$) for all $\alpha$. On the other hand, the stochastic thermostat model leads to a coefficient that is negative and small for weak inelasticity, becoming positive for $\alpha \alt 0.7$, and increasing very slowly afterwards \cite{PTvNyE01}. The fair conclusion is that none of the models is able to reproduce the qualitative behavior of the Sonine coefficient, but both accurately predict that the coefficient is very small in the quasi-elastic regime. To put this conclusion in a proper context, it is important to stress that the theoretical predictions for $a_{2,st}$ have been both obtained by using a linear approximation in which only terms to first order in $a_{2}$ are kept in the calculations \cite{PTvNyE01,BGMyB13}.

\begin{figure}
\includegraphics[scale=0.4,angle=0]{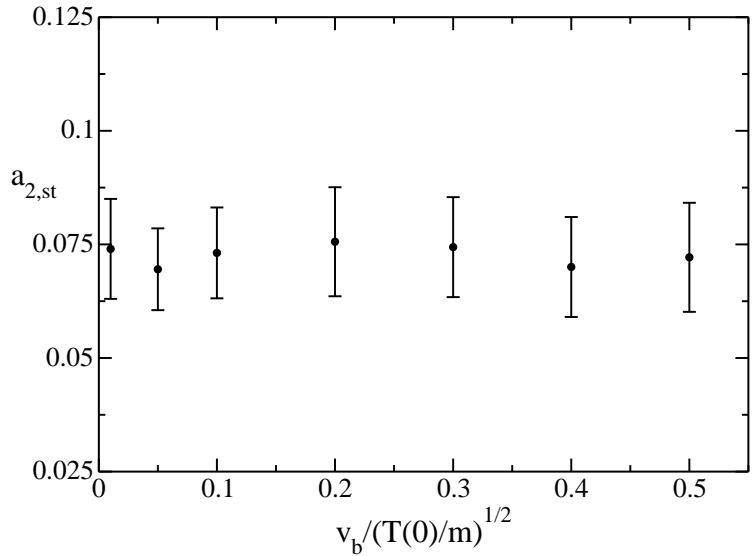}
\caption{ MD simulation results for the Sonine coefficient $a_{2,st}$ of the steady velocity distribution function of the quasi-two-dimensional granular gas  as a function of the dimensionless velocity of the vibrating wall. The coefficient of normal restitution is $\alpha= 0.8$, the density of the three-dimensional gas is $n=0.02 \sigma^{-3}$, and the distance between the two confining walls is $h= 1.5 \sigma$. \label{f6}}
\end{figure}

The dependence of  $a_{2,st}$ on $v_{b}$ is analyzed in Fig.\ \ref{f6}, for a system with $\alpha=0.8$ and the same values of the other parameters as in Fig.\,   \ref{f5}. No dependence on $v_{b}$ is observed. The same happens for other values of the restitution coefficient. This is consistent with the predictions following from both the stochastic and the collisional models, in which the steady Sonine coefficient is independent from $\xi_{0}$ and $\Delta$, respectively. 

\begin{figure}
 \includegraphics[scale=0.4,angle=0]{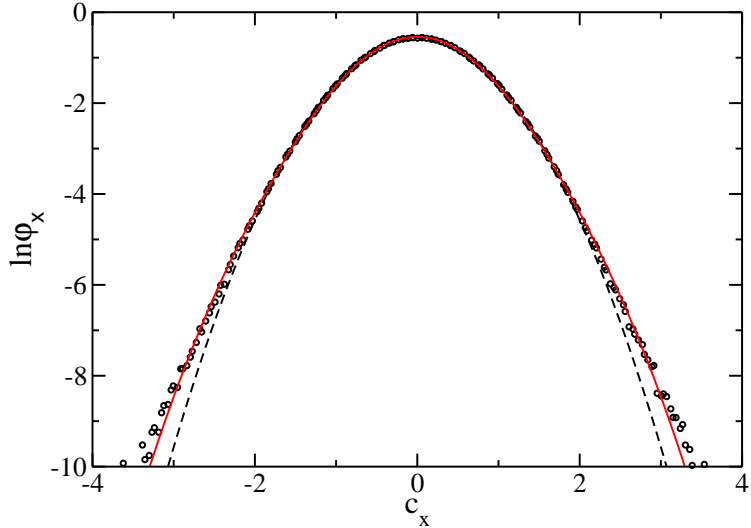}
\caption{ (Color online) Logarithmic  plot of the MD simulation values (black circles) of the marginal one-particle velocity distribution $\varphi_{x}(c_{x})$ defined by Eq. (\ref{3.1}).  The (black) dashed line is the Gaussian and the (red) solid line is the Sonine approximation as given by Eq.  (\ref{3.3}) with the value of $a_{2,st}$ measured also in the simulations.  The parameters of the system are the same as described in the legend of Fig. \protect{\ref{f6}}. \label{f7}}
\end{figure}

The fact that the Sonine coefficient $a_{2}$ is small does not guarantee by itself that the one-particle distribution function can be accurately approximated by a Gaussian. It could happen that contributions of higher order cumulants be relevant and even that the Sonine expansion be divergent. To see whether this is the case, in Fig. \ref{f7} the marginal one-particle distribution function $\varphi_{x} (c_{x})$ measured by MD simulations is shown, again  for $\alpha=0.8$, $n= 0.02 \sigma^{-3}$, and $h=1.5 \sigma$. 
Also plotted are the Gaussian and the first Sonine approximation, Eq. (\ref{3.3}), using the value of $a_{2,st}$ measured in the MD  simulations. The agreement between the Sonine approximation and the actual velocity distribution is very good. Also notice that the deviation of the distribution from the Maxwell-Boltzmann one is clearly observed. In summary, the velocity distribution describing the horizontal motion of the particles is very well represented by Eq.\ (\ref{3.3}), at least in thermal region, i.e. for values of $c$ of the order of unity. This is similar to what has been found for the thermostated model \cite{MyS00} and also for  the colisional one \cite{BGMyB13}.

\section{The Kovacs-like effect in the collisional model}
\label{s4}

In the context of kinetic theory, the existence of a macroscopic hydrodynamic regime, in which the time evolution of the temperature obeys a closed equation, requires that the one-particle distribution function of the system be {\em normal}, meaning that all its time dependence occurs through the temperature \cite{RydL77,DyB11}. Then, dimensional analysis leads to  a scaling of the velocity distribution of the collisional model of the form
\begin{equation}
\label{4.1}
f_{H}({\bm v},t)= n v_{0}(t)^{-2} \varphi ({\bm c}, \Delta^{*}),
\end{equation}
where $v_{0}(t)$ and ${\bm c}$ are defined below Eq.\  (\ref{2.13}) and in Eq.\ (\ref{2.15}), respectively.  The index $H$ indicates that we are restricting ourselves to homogenous hydrodynamic states. Consequently, if the first Sonine approximation (\ref{2.14}) is employed, it follows that the dimensionless Sonine coefficient $a_{2}$ must also be normal, i.e. 
\begin{equation}
\label{4.2}
a_{2}(t) \rightarrow \overline{a}_{2}(t)= \overline{a}_{2} \left( \frac{T(t)}{T_{st}} \right).
\end{equation}
Equivalently, the time-dependent temperature could have been scaled with $m \Delta^{2}$. The existence of a normal solution in which $a_{2}(t)$ is of the above form and, therefore, of homogeneous hydrodynamics, has been established for both the stochastic model \cite{GMyT12} and  the collisional model \cite{BMGyB14}. 

\begin{figure}
\includegraphics[scale=0.4,angle=0]{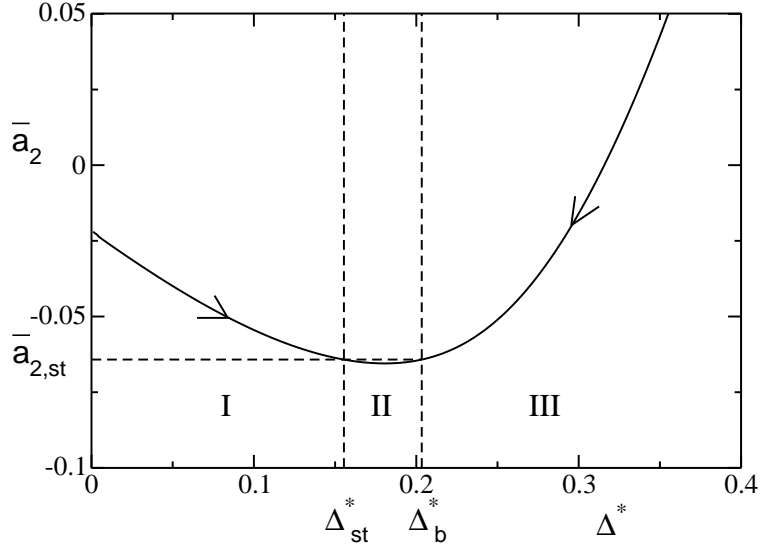}
\caption{ Hydrodynamic Sonine coefficient $\overline{a}_{2}$, as a function of the dimensionless characteristic speed $\Delta^{*}$ for a system with a coefficient of normal restitution $\alpha=0.8$. Shown in the figure are the steady values of $a_{2}$  and  $\Delta^{*}$,   $a_{2,st}$  and  $\Delta^{*}_{st} $, respectively. Also plotted is the value $\Delta^{*}_{b}$, defined by Eq.\ (\protect{\ref{4.16}}). The arrows on the curve indicate the way in which it is described as the system approaches its steady state. For constant $\Delta$, interval  $I$ corresponds to cooling and intervals $II$ and $III$ to heating. If the value of $a_{2}$ at the moment $t=t_{0}$ in which $\Delta$ is modified so that $\Delta^{*}(t_{0})=\Delta^{*}_{st}$, lies in regions $I$ or $II$, the system keeps on cooling and heating, respectively. On the other hand, if it lies in region $III$ the system instantaneously changes from heating to cooling (rebound effect).              \label{f10}}
\end{figure}

Suppose a two-dimensional system of hard disks with the collision rule given by Eqs.\ (\ref{2.10}) and (\ref{2.11}) and  with a characteristic speed $\Delta= \Delta_{1}$.  The system is in the steady state, so that its temperature, $T_{1,st}$, is given by
\begin{equation}
\label{4.3}
\frac{\Delta_{1}}{v_{01,st}}= \Delta^{*}_{st},
\end{equation}
where 
\begin{equation}
\label{4.4}
v_{01,st} \equiv \left(\frac{ 2 T_{1,st}}{m} \right)^{1/2}
\end{equation}
and $\Delta^{*}_{st}$ fulfils the equation (see Eq. (\ref{2.12}))
\begin{equation}
\label{4.5}
\zeta^{*}(\alpha, \Delta^{*}_{st})=0.
\end{equation}
Here, the dimensionless hydrodynamic rate of variation of the energy,
\begin{equation}
\label{4.6}
\zeta^{*}(\alpha,\Delta^{*}) \equiv ( 2\pi)^{1/2} \left[ \frac{1 - \alpha^{2}}{2} \left( 1+ \frac{3 \overline{a}_{2}(\Delta^{*}) }{16} \right) - \alpha \left( \frac{\pi}{2} \right) ^{1/2} \Delta^{*} - \left( 1- \frac{\overline{a}_{2}(\Delta^{*})}{16} \right) \Delta^{*2} \right]
\end{equation}
has been introduced. Then, at a given moment, the characteristic speed is changed to $\Delta_{2} < \Delta_{1}$. The steady temperature $T_{2,st}$ corresponding to the new value of the characteristic speed will be given by
\begin{equation}
\label{4.7}
\left(\frac{ 2 T_{2,st}}{m} \right)^{1/2} = \frac{\Delta_{2}}{\Delta^{*}_{st}} < \frac{\Delta_{1}}{\Delta^{*}_{st}} = \left(\frac{ 2 T_{1,st}}{m} \right)^{1/2}.
\end{equation}
Consequently, the system will monotonically cool towards $T_{2,st}$, at least after some non-hydrodynamic initial transient. Let us denote the time dependent temperature once in the hydrodynamic regime by $T(t)$. The one-particle distribution function of the system describing its time  evolution in the first Sonine approximation is given by Eq.\ (\ref{2.14}) with
\begin{equation}
\label{4.8}
a_{2}(t) = \overline{a}_{2} ( \alpha, \Delta^{*}_{2}(t)),
\end{equation}
\begin{equation}
\label{4.9}
\Delta^{*}_{2}(t) \equiv \frac{\Delta_{2}}{v_{0}(t)}.
\end{equation}
Consider next a later time $t_{0}$, and be $T_{0}=T(t_{0})$ the temperature of the system at that time. The cooling rate at the same  moment is
\begin{equation}
\label{4.10}
\zeta(\alpha, \Delta^{*}_{2}(t_{0}))= n \sigma v_{0}(t_{0}) \zeta^{*} (\alpha, \Delta^{*}_{2}(t_{0})).
\end{equation}
As indicated above, the system is cooling and, therefore, it is
\begin{equation}
\label{4.11}
\zeta^{*}(\alpha, \Delta^{*}_{2} (t_{0})) >0.
\end{equation}
There is a value $\Delta_{0}$ of the characteristic speed for which
\begin{equation}
\label{4.12}
\frac{\Delta_{0}}{v_{0}(t_{0})} = \Delta^{*}_{st},
\end{equation}
so that $T_{0}$ is precisely the steady temperature for $\Delta= \Delta_{0}$. In the  experiment being described, the value of the characteristic speed is instantaneously modified at $t=t_{0}$ to the value $\Delta_{0}$, defined above. Therefore, the cooling rate of the system changes to
\begin{equation}
\label{4.13}
\zeta_{+}(\alpha, t_{0} )= n \sigma v_{0}(t_{0}) \zeta^{*}_{+} (\alpha, t_{0}),
\end{equation}
with
\begin{eqnarray}
\label{4.14}
\zeta^{*}_{+}(\alpha,t_{0})  &= & ( 2\pi)^{1/2} \left[ \frac{1 - \alpha^{2}}{2} \left( 1+ \frac{3 \overline{a}_{2}(\Delta^{*}_{2}(t_{0})) }{16} \right) - \alpha \left( \frac{\pi}{2} \right) ^{1/2} \Delta^{*}_{st} \right. \nonumber \\ 
&  &  \left.- \left( 1- \frac{\overline{a}_{2}(\Delta^{*}_{2}(t_{0}))}{16} \right) \Delta^{*2}_{st} \right].
\end{eqnarray}
Using Eq. (\ref{4.5}), this expression can be rewritten as
\begin{equation}
\label{4.15}
\zeta^{*}_{+}(\alpha, t_{0}) = \frac{( 2\pi)^{1/2}}{16}  \left[  \frac{3(1-\alpha^{2})}{2}+ \Delta_{st}^{*2} \right] \left[ \overline{a}_{2} (\Delta_{2}^{*}(t_{0}))- \overline{a}_{2} \left( \Delta^{*}_{st} \right) \right].
\end{equation}
It follows that the cooling rate $\zeta_{+}(\alpha, t_{0})$ is positive if $\overline{a}_{2} (\Delta_{2}^{*}(t_{0})) >\overline{a}_{2}(\Delta^{*}_{st})$, and negative if $\overline{a}_{2} (\Delta_{2}^{*}(t_{0})) <\overline{a}_{2}(\Delta^{*}_{st})$. It only vanishes if  $\overline{a}_{2} (\Delta_{2}^{*}(t_{0})) = \overline{a}_{2}(\Delta^{*}_{st})$, i.e. the system was at $t=t_{0}$ in the steady state. In Fig. \ref{f10}, the hydrodynamic Sonine coefficient $\overline{a}_{2}$ is plotted as a function of $\Delta^{*}$  for a system with $\alpha=0.8$ \cite{BMGyB14}. The steady value $\Delta_{st}^{*}$ is indicated. In a cooling experiment, as the one ocurring when $\Delta$ is changed to $\Delta_{2}$,  the value of $\Delta^{*}$ is increasing towards its steady value, and $\overline{a}_{2}$ is decreasing towards $\overline{a}_{2,st}$ . It follows that  it is always $\overline{a}_{2} (\Delta_{2}^{*}(t_{0})) >\overline{a}_{2}(\Delta^{*}_{st})$. This conclusion holds as long as the dependence of $\overline{a}_{2}$ on $\Delta^{*}$ be similar to the one illustrated in Fig.\ \ref{f10}. Actually, this happens in the collisional model for all the physical values of the restitution coefficient, $ 0 < \alpha \leq 1$.

Consider next the same experiment, but this time being $\Delta_{1} <\Delta_{2}$, i.e. the system is increasing monotonically its temperature when $\Delta$ is instantaneously changed  from $\Delta_{2}$ to $\Delta_{0}$ at $t=t_{0}$. In other words, its cooling rate is negative just before the change. What happens just after the change? If $\overline{a}_{2}(\Delta^{*}(t_{0})) >\overline{a}_{2}(\Delta^{*}_{st})$, the cooling rate is positive after the change and the system is cooling; its temperature decreases in time. On the other hand, if $\overline{a}_{2}(\Delta^{*}(t_{0})) <\overline{a}_{2}(\Delta^{*}_{st})$, the system will keep on heating. The former case can be understood as a rebound effect in the evolution of the temperature. The above discussion can expressed alternatively as follows. Define $\Delta^{*}_{b}$ as the value $\Delta^{*}>\Delta_{st}^{*}$  verifying
\begin{equation}
\label{4.16}
\overline{a}_{2} (\Delta^{*} _{b}) = \overline{a}_{2,st}.
\end{equation}
If $\Delta^{*}_{2}(t_{0}) >\Delta^{*}_{b}$, there is rebound  effect of the temperature, while if $\Delta^{*}_{st} < \Delta^{*}_{2}(t_{0}) <\Delta^{*}_{b}$, there is no such effect, and the temperature continues increasing after the modification of $\Delta$. In Fig. \ref{f10}, the several intervals  for $a_{2}(\Delta^{*}_{2}(t_{0})) $, defining the qualitative behavior of the system after modification of the characteristic speed $\Delta$ are indicated.

The fact that the temperature of the system after the change of the characteristic speed does not remain in its steady value, but deviates initially   from it to decay afterwards, can be considered as reminiscent of the Kovacs effect exhibited by molecular system in their equilibrium relaxation \cite{Ko63}. Evidently, this happens because the system is not in the steady state, in spite of having the steady temperature. In the context of kinetic theory, this implies that its velocity distribution function is not fully determined by the instantaneous temperature, i.e. it is not {\em normal}. 

\begin{figure}
\includegraphics[scale=0.4,angle=0]{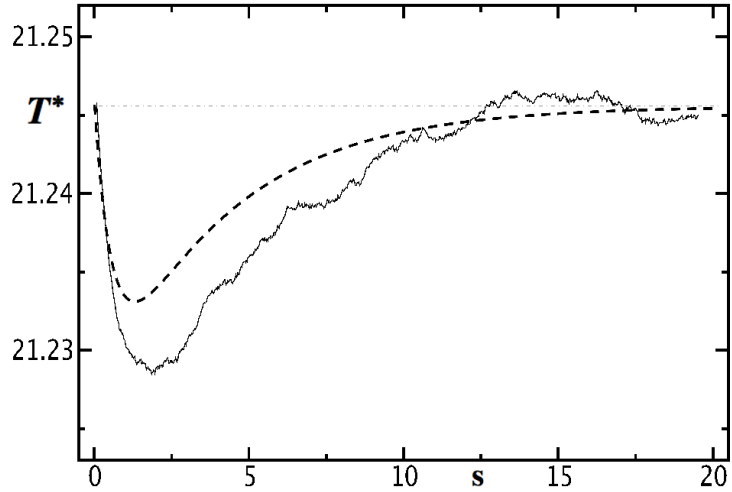}
\caption{ The Kovacs effect in a system under cooling. Temperature $T^{*}$ and time $s$ are measured in the dimensionless units indicated in the main text. The characteristic speeds verify $\Delta_{2}=0.5 \Delta_{0}$, where $\Delta_{2}$ is  the speed in the former (no plotted) evolution, and $\Delta_{0}$ is  the speed in the final relaxation starting at the steady temperature.  The dashed line is the theoretical prediction given by the solution of Eqs. (\protect{\ref{ap1.1}}) and (\protect{\ref{ap1.2}}),  while the solid line has been obtained by the DSMC method. \label{f11}}
\end{figure}

In the Sonine approximation for the one-particle velocity distribution given by Eq. (\ref{2.14}), the time evolution of the temperature in an homogeneous system can be described by means of a pair of coupled differential equations. Assuming that the Sonine coefficient $a_{2}$ and its time  derivative are small, the equations can be written in the form given in the Appendix, Eqs. (\ref{ap1.1}) and (\ref{ap1.2}). In the following, numerical solutions of those equations will be reported in order to verify and also quantify the conclusions reached in the discussion carried out above. In Fig. \ref{f11} the dimensionless temperature $T^{*}(t) \equiv T(t) /m \Delta^{2}$ is plotted as a function of the dimensionless time $s$ given by $ds  \equiv v_{0}(t) n \sigma dt$, where $v_{0}$ is the thermal velocity defined  above Eq. (\ref{2.14}). The time scale $s$ is proportional to the accumulated number of collisions per particle in the system, and the origin $s=0$ has been chosen for $t=t_{0}$, so that only the evolution after putting $\Delta = \Delta_{0}$ is plotted.  The values of the characteristic speeds  of the process are related by  $\Delta_{0}= 2 \Delta_{2}$. This relation fully specify the solution for $t > t_{0}$, as discussed in the Appendix.  Since $\Delta_{0} > \Delta_{2}$, the system was cooling before the change of the characteristic speed to $\Delta_{0}$. As predicted by the above discussion, the system keeps cooling after the change, approaching later on its steady value from below. 

Since the model predictions for the Sonine coefficient $a_{2}$ qualitatively disagree with the MD results, it is clear that the effect being analyzed now must be different in the model and in the original quasi-two-dimensional confined gas. For this reason, it seems more illuminating to consider the numerical results obtained by the direct simulation Monte-Carlo method (DSMC) applied to the collisional model than MD results. The DSMC method is a particle simulation algorithm designed to mimic the dynamics of a low density gas \cite{Bi94,Ga00}, i.e. a gas obeying the Boltzmann equation. This simulation method has been extensively applied to granular gases, and it will not be described here. Let us only mention that the homogeneity of the system being simulated can be exploited to improve the accuracy of the results \cite{BRyC96}. The number of particles employed in the simulations is $N=1000$ and the reported results have been averaged over $ 5 \times 10^{4}$ trajectories.  Included in Fig.\ \ref{f11} are the DSMC results for a system with $\Delta_{2}= 0.5 \left( T(0) 
/m \right)^{1/2}$ and $\Delta_{0}= \left( T(0) /m \right)^{1/2}$, where $T(0)$ is a reference initial temperature. The agreement between the simulation results and the theoretical prediction can be considered as satisfactory, taking into account the approximations involved in the derivation of Eqs. (\ref{ap1.1}) and (\ref{ap1.2}). Note that the plotted curves start at $t=t_{0}$ ($s=0$), at which the characteristic speed is changed from $\Delta_{2}$ to $\Delta_{0}$.

\begin{figure}
 \includegraphics[scale=0.4,angle=0]{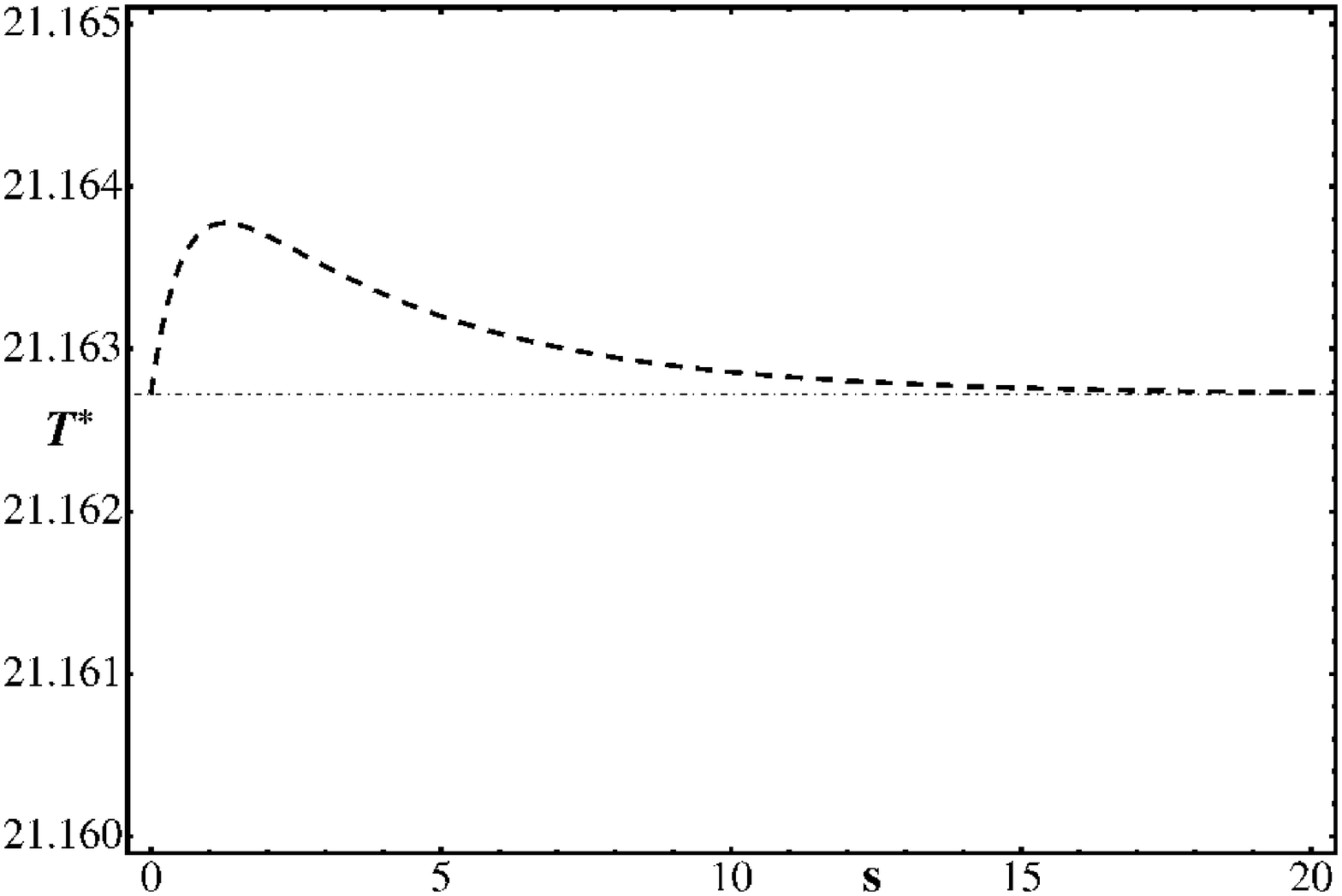}
\caption{ The Kovacs effect without rebound in a system under heating. Temperature $T^{*}$ and time $s$ are measured in the dimensionless units indicated in the main text. The characteristic speeds verify $\Delta_{2}= 1.2  \Delta_{0}$, where  $\Delta_{2}$ is  the speed in the former evolution, and $\Delta_{0}$ is  the speed in the final relaxation starting at its steady temperature.  The dashed curve is the solution of the evolution equations given in the Appendix. \label{f12}}
\end{figure}

\begin{figure}
 \includegraphics[scale=0.4,angle=0]{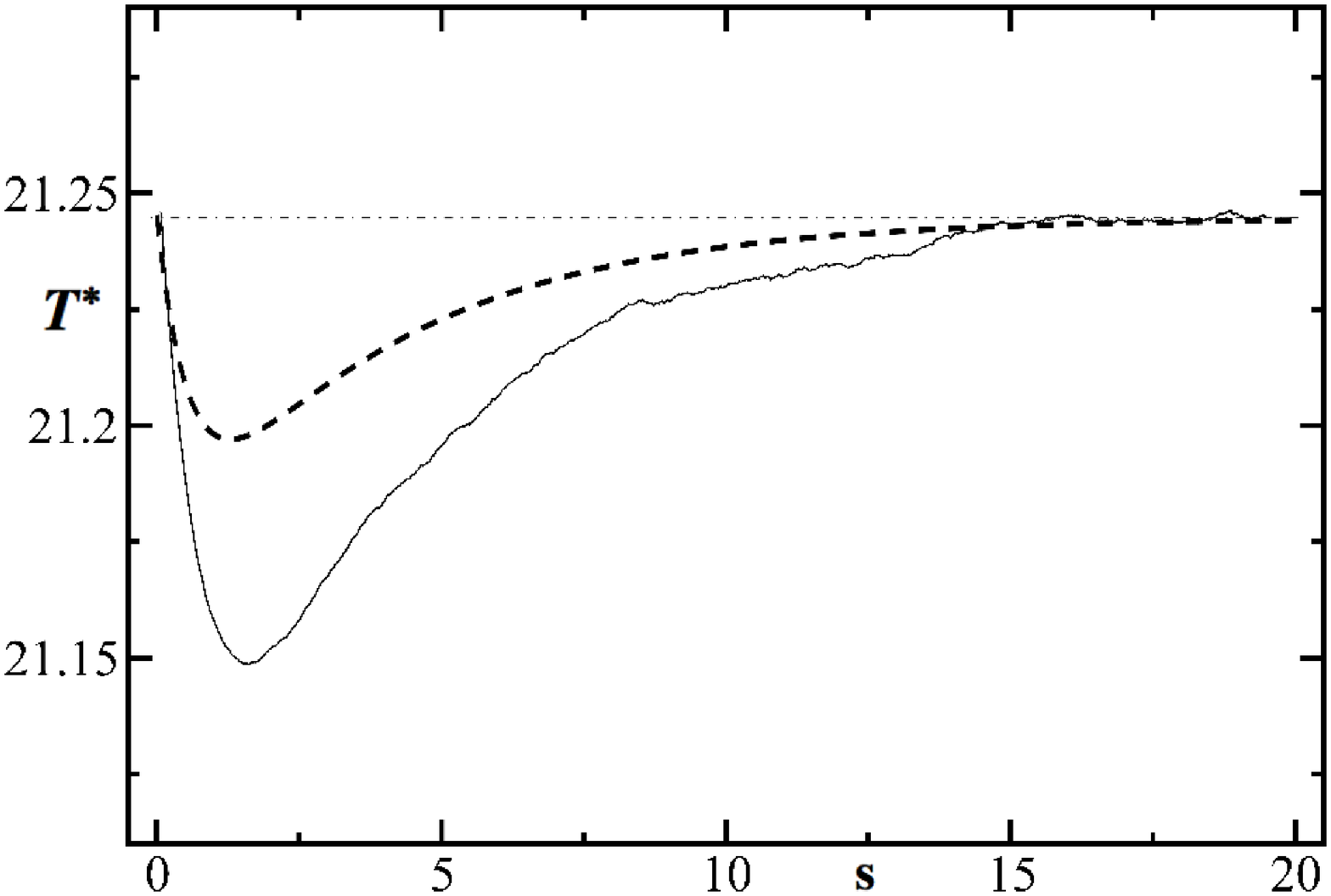}
\caption{ The Kovacs effect with rebound in a system under heating. Temperature $T^{*}$ and time $s$ are measured in the dimensionless units indicated in the main text. The characteristic speeds verify $\Delta_{2}= 2 \Delta_{0}$, where  $\Delta_{2}$ is  the speed in the former evolution, and $\Delta_{0}$ is  the speed in the final relaxation starting at its steady temperature.  The dashed line is the theoretical prediction given by the solution of Eqs. (\protect{\ref{ap1.1}}) and (\protect{\ref{ap1.2}}),  while the solid line has been obtained by the DSMC method.  Note that the temperature of the system was increasing for $s<0$. \label{f13}}
\end{figure}

The behavior of the system in heating experiments is illustrated in Figs \ref{f12} and \ref{f13}. The former is for $\Delta_{2}= 1.2 \Delta_{0}$, while the latter is for $\Delta_{2} = 2 \Delta_{0}$. In both cases the system is heating  just before $\Delta$ is modified from $\Delta_{2}$ to $\Delta_{0}$. The difference between  them is that the value of $a_{2}(t_{0})= \overline{a}_2 \left( \Delta_{2} \Delta^{*}_{st} / \Delta_{0} \right) $ at that time is in the first case smaller than $\overline{a}_{2,st}$, while  it is larger in the second one. In agreement with the discussion carried out above, the temperature keeps on increasing just after the change of the peculiar velocity to $\Delta_{0}$ in the first experiment, while it decreases (rebound effect) in the  second one. Note that only the evolution for $t>t_{0}$ is plotted.  DSMC simulation results are shown only for one of the cases, namely for $\Delta_{2}= 2 \left( T(0) )/m \right)^{1/2}$ and $\Delta_{0}= \left( T(0) / m \right)^{1/2}$.  This is because the values of $\overline{a}_{2} (\Delta^{*})$ for $ \Delta^{*}_{st} < \Delta^{*} < \Delta^{*}_{b}$ are very close to $\overline{a}_{2,st}$, as it is seen in Fig.\ \ref{f10}. As a consequence, when starting with a value of $a_{2}$ within the interval $II$ in Fig.\ \ref{f10}, the effect is very weak, and hard to measure in the DSMC simulations because of the statistical uncertainties. The different scales used on the vertical axis of Figs. \ref{f12} and \ref{f13} must be stressed.

Similar results have been obtained for other relations of the charateristic speeds and also for $\alpha=0.9$, then expecting the same degree of agreement between theory and DSMC simulations over a wide range of parameters. 

\section{Concluding remarks}
\label{s5}
In the first part of this paper, the accuracy of the collisional model introduced by Brito {\em et al.} \cite{BRyS13} to describe the homogeneous two-dimensional dynamics of a confined gas  of inelastic hard spheres has been investigated. The phenomenological speed parameter involved in the formulation of the model has been determined by fitting the MD simulation values of the steady temperature.  Next, the time evolution of the temperature when relaxing towards the steady state predicted by the model using that value has been compared with the simulation results, and an acceptable good agreement has been found. Moreover, the fitted characteristic speed of the model is proportional to the vibration velocity of the walls in the original system, what is consistent with its physical interpretation. A similar program has been carried out with a thermostated random model, often used in the literature to generate a steady state of granular system \cite{vNyE98,PTvNyE01,GyM02,VPBTyvW06,GMyT09,MGyT09}, getting  a worse agreement with the simulation data. Also the one-particle velocity distribution function of the system in the steady state has been studied. In the simulations, as well as in the two models considered, the steady distribution turns out to be quite accurately well described by a Gaussian plus a small correction given by the first Sonine approximation. The coefficient measuring the magnitude of the correction term only depends on the value of the coefficient of normal  restitution, but not of the parameter measuring the rate of injection of energy in the system.

The first Sonine approximation also describes accurately the homogeneous hydrodynamics of the collisional model \cite{BMGyB14}. In this context, it must be emphasized that, although the Sonine coefficient be quantitatively small, its existence and dependence on time through the temperature is essential for the possibility of a hydrodynamic description. The second part of the paper deals with a memory effect in the time evolution of the temperature in the collisional model that is also well captured by the first Sonine approximation. The characteristic speed parameter is suddenly changed while the system is relaxing towards its steady state, so that the latter is modified. This is done  in such a way that the system is instantaneously at the steady temperature corresponding to the new speed. Then, the posterior evolution of the temperature exhibits a highly non-monotonic behavior in which the temperature of the system initially deviates from its long time steady value. This  behavior somehow resembles  the Kovacs effect occurring in the relaxation to equilibrium of molecular systems \cite{Ko63}.  The analysis presented here largely benefits from considering that, when the characteristic speed is modified, the system was evolving accordingly with the hydrodynamic law. This allows to determine its distribution function at that moment just from the knowledge of the temperature.  Of course, afterwards this property is lost, the distribution function is not determined by the temperature (i.e. it is not {\em normal}) and the system is not in the hydrodynamic regime. 

Under the above hydrodynamic initial condition, it is easy to explain the existence of memory effects in the collisional model and to predict the qualitative behavior of the temperature of the system after the jump of the  characteristic speed, i.e. whether the temperature keeps on varying monotonically or there is some kind of rebound effect, with the temperature rate changing sign at the moment of the jump. Moreover, quantitative predictions are given by the (numerical) solutions of two coupled first order differential equations. A good agreement has been found between the theoretical predictions and simulation values obtained by the direct simulation Monte Carlo method.

The reported results show that highly non-linear and non-hydrodynamic evolution of a system can be quite easily understood by means of simple kinetic theory arguments. Quite peculiarly, here this has been done by taking advantage of choosing the initial evolution state as being described by hydrodynamics. Moreover, the analysis suggests the possibility of experimentally observing the nonlinear relaxation of the Kovacs-like effect in a purely mechanical system, namely the confined quasi-two-dimensional granular gas submitted to vertical vibration. The restriction to initial hydrodynamic states seems something easily implementable. Then, on the basis of the discussion in the first part of the paper, one could expect that the general trends obtained with the collisional model remain valid in the real system. For instance, it would not be surprising that the system showed rebound effect (practically) always if the jump is done while it is heating and never if done while it is cooling.

\section{Acknowledgements}

This research was supported by the Ministerio de Educaci\'{o}n y Ciencia (Spain) through Grant No. FIS2011-24460 (partially financed by FEDER funds).

\appendix*

\section{Time evolution of the temperature in the Sonine approximation}
\label{ap1}
By taking moments in the Boltzmann equation and considering the Sonine approximation given by Eq.\ (\ref{2.14}), the following pair of coupled first order differential equations are obtained:
\begin{equation}
\label{ap1.1}
\frac{\partial \Delta^{*}}{\partial \tau} = \frac{(2 \pi)^{1/2}}{8}  \left[ A_{0}+A_{1} a_{2}(t) \right],
\end{equation}
\begin{equation}
\label{ap1.2}
\frac{\partial a_{2}}{\partial \tau}= \frac{(2 \pi)^{1/2}}{8 \Delta^{*}(t)} \left\{ \left[ 4 A_{0}+ B_{0} \right] + \left[ 4(A_{0}+A_{1}) +B_{1} \right] a_{2}(t) \right\},
\end{equation}
\begin{equation}
\label{ap1.3}
A_{0} (\alpha, \Delta^{*})= 4 \left[ \frac{1- \alpha^{2}}{2} - \left( \frac{\pi}{2} \right)^{1/2}  \alpha \Delta^{*} - \Delta^{*2} \right],
\end{equation}
\begin{equation}
\label{ap1.4}
A_{1} (\alpha, \Delta^{*}) = \frac{1}{4} \left[ \frac{3(1- \alpha^{2})}{2}\, + \Delta^{*2} \right],
\end{equation}
\begin{eqnarray}
\label{ap1.5}
B_{0}(\alpha, \Delta^{*}) &=& (2 \pi )^{1/2} \left(5+3 \alpha^{2} +4 \Delta^{*2}\right) \alpha \Delta^{*}-3+4 \Delta^{*4} + \alpha^{2}+ 2 \alpha^{4}  \nonumber  \\
&& -4 \left( 1-\alpha^{2}-2 \Delta^{*2} \right)+ 2 \Delta^{*2} \left( 1+6 \alpha^{2} \right),
\end{eqnarray}
\begin{eqnarray}
\label{ap1.6}
B_{1}(\alpha,\Delta^{*})& = & \left( \frac{\pi}{2} \right)^{1/2} \left[ 2 -4(1-\alpha)+7 \alpha+3 \alpha^{3} \right] \Delta^{*}- \frac{1}{16} \left\{ 85 + 4 \Delta^{*4}-18 (3+2 \alpha^{2}) \Delta^{*2} \right.  \nonumber \\
&& \left. - \left( 32 +87 \alpha+ 30 \alpha^{3} \right) \alpha -4 \left[ 6 \Delta^{*2}-(1+\alpha) (31-15 \alpha) \right] \right\}.
\end{eqnarray}
In the above expressions, a dimensionless time scale,
\begin{equation}
\label{ap1.7}
\tau \equiv \Delta n \sigma t,
\end{equation}
has been introduced. The general method leading to Eqs. (\ref{ap1.1}) and (\ref{ap1.2}) has been discussed in Ref. \cite{BMGyB14} and it will be not repeated in detail here.  Let us only indicate that the former equation is obtained by multiplying the Boltzmann equation by the square of the velocity of the particles, integrating afterwards over it, while for the latter equation, one has to multiply the Boltzmann equation by the fourth power  of the velocity before integrating. Moreover, upon deriving Eq. (\ref{ap1.2}) terms quadratic in $a_{2}$  have been neglected.

All the possible dependence of the solutions of Eqs.\ (\ref{ap1.1}) and (\ref{ap1.2}) on the characteristic speeds $\Delta_{2}$ and $\Delta_{0}$ occurs through the initial conditions. Let us focus on the evolution of the system after $\Delta$ is instantaneously changed from $\Delta_{2}$ to $\Delta_{0}$ at $t=t_{0}$. The fact that the system is in the hydrodynamic regime just before the change, and the way in which  $\Delta_{0}$ is chosen have a relevant consequence. The initial conditions to be considered are given by
\begin{equation}
\label{ap1.8}
\Delta^{*}(t=t_{0})= \Delta^{*}_{st},
\end{equation}
\begin{equation}
\label{ap.9}
a_{2}(t=t_{0})= \overline{a}_{2} \left[ \Delta^{*}_{2}(t_{0}) \right] = \overline{a}_{2}  \left( \frac{\Delta_{2}}{\Delta_{0}}\, \Delta^{*}_{st} \right).
\end{equation}
It follows that the solution, and therefore the evolution of the temperature,  is fully specified by the ratio of the characteristic speeds $\Delta_{0}/\Delta_{2}$, being independent of each of them separately. A similar argument can be applied to the previous evolution of $\Delta^{*}$ with $\Delta= \Delta_{2}$.

\end{document}